\documentclass[letterpaper,titlepage,12pt]{article}
\usepackage[colorlinks=true,urlcolor=blue,citecolor=magenta]{hyperref}
\usepackage{amssymb,amsmath,amsfonts}
\usepackage{epsfig}
\usepackage{epsfig}
\usepackage{graphicx}
\usepackage{epstopdf}
\usepackage[round]{natbib}
\usepackage{caption}
\usepackage{subcaption}
\usepackage{amscd}
\usepackage{amsthm}
\usepackage{csquotes}
\usepackage{latexsym}
\usepackage{float}
\usepackage{mathtools}
\usepackage[stable]{footmisc}
\usepackage{amsbsy}
\usepackage{bbm}
\usepackage[english]{babel}
\usepackage{psfrag}
\usepackage{tabularx}
\usepackage{colortbl}
\usepackage{xcolor}
\usepackage{braket}

\allowdisplaybreaks

\newcommand{\be}{\begin{eqnarray}}
\newcommand{\ee}{\end{eqnarray}}
\newcommand{\beq}{\begin{eqnarray}}
\newcommand{\eeq}{\end{eqnarray}}

\def\clock{{\count0=\time
  \divide\count0 60
  \ifnum\count0<10 0\fi\the\count0
  \multiply\count0 -60 \advance\count0 \time
  :\ifnum\count0<10 0\fi \the\count0
  }}
\newcommand{\timestamp}{{\small\vbox{\hbox{\tt\jobname.tex}
\hbox{\the\day/\the\month/\the\year, \clock}}}}

\numberwithin{equation}{section}

\begin{document}

\vskip 1.4 cm
\centerline{\LARGE \bf Humeanism in Light of Quantum Gravity}
\vskip 1.5cm
\centerline{\large {{\bf Enrico Cinti$^{1,2}$, Marco Sanchioni$^1$}}}
\vskip .8cm
\begin{center}
\sl $^1$ DISPeA, University of Urbino,\\
\sl Via Timoteo Viti 10, 61029 Urbino PU , Italy.\\
\end{center}
\begin{center}
\sl $^2$ Department of Philosophy, University of Gen\`eva,\\
\sl 5, rue de Candolle, CH-1211 Gen\`eve 4, Switzerland.\\
\end{center}
\vskip 0.6cm
\centerline{\small\tt marco.sanchioni2@gmail.com, cinti.enrico@gmail.com}
\vskip .8cm

\centerline{ \textit{Forthcoming in Synthese}}

\vskip .8cm \centerline{\bf Abstract} \vskip 0.1cm 

\noindent Quantum Theory and Humeanism have long been thought to be incompatible due to the irreducibility of the correlations involved in entangled states. In this paper, we reconstruct the tension between Humeanism and entanglement via the concept of causal structure, and provide a philosophical introduction to the ER=EPR conjecture. With these tools, we then show how the concept of causal structure and the ER=EPR conjecture allow us to resolve the conflict between Humeanism and entanglement. 


\tableofcontents

\section{Introduction}\label{1}
The conflict between the doctrine of Humeanism and quantum entanglement has historically been central to metaphysical foundations of quantum mechanics (QM). In particular, many different proposals have flourished with the goal of either finding alternative formulations of QM which remove the problematic features of entanglement \citep{10.1093/bjps/axt019, Esfeld2014-ESFQHO} or of individuating variants of the Humean thesis which are compatible with QM despite entanglement \citep{Lyre2010, Darby2012-DARRHA, Miller2014-MILQEB, 10.2307/24704648, Bhogal2017-BHOWTH}. The aim of our paper is different from both these approaches. Our approach falls squarely in the context of \textit{entanglement realism} (as understood and defended against more eliminative approaches in \cite{pittphilsci14945}), roughly speaking the view that there are relations constitutive of entanglement in our fundamental ontology. Indeed, one might see this work as a specific proposal for a form of entanglement realism and its compatibility with Humeanism, in the context of quantum gravity (QG). We thus argue that one can attack the problem of reconciling entanglement and Humeanism not by modifying either one or the other, but instead by looking at some tantalising approaches and conjectures regarding the extension of QM to the realm of gravity and spacetime. Studying the interaction of QG with metaphysics is a practice more and more common in the literature, as witnessed by \cite{le2019spacetime, Matarese:2019bxo,pittphilsci15902, pittphilsci17099}. This approach holds much promise for the interaction between the edge of physical and metaphysical research. \\ 
Before getting to the core of our discussion, let us briefly address a preliminary question: why should QG be relevant to this problem? Or, more generally, why should QM not be enough to address the issue of reconciling entanglement with Humeanism? There are two sets of reasons for extending our sight beyond the realm of QM to QG: on the one hand, one might note that up to this point, most resolutions of the conflict between QM and Humeanism have proven unsatisfactory. Indeed, the fact that the debate is still going on shows that there are no universally accepted answers to these problems. Such a situation suggests that there might be something to the idea of trying new avenues to find possible alternatives to the standard options in the debate. On the other hand, and a more substantive note, QG seems to be especially relevant to the question of the relationship between entanglement and Humeanism, since this tension, as we make clear in the next sections, ultimately rests on the conflict between the causal structure of spacetime and the relations of entanglement. Since QG should be a theory which deals with quantum spacetime, and thus directly with the interplay between spacetime structure and entanglement, it is uniquely situated to provide insight on the compatibility of Humeanism and entanglement. Indeed, having a quantum spacetime is the goal of most theories of QG. Identifying ways in which these two sets of structures can be compatible is one of its main tasks. \\
Specifically, in this paper, we focus on the so-called ER=EPR conjecture \citep{Maldacena:2013xja}, a speculative extension of the picture of spacetime coming from General Relativity (GR). ER=EPR incorporates some initial quantum effects into classical spacetime structure. As we argue in \S\ref{5}, it is a reasonable possibility that such conjecture is satisfied in the semiclassical regime of a complete theory of QG. For this reason, while ER=EPR is a conjecture, we think it can be useful to understand how this conjecture might impact the compatibility between Humeanism and entanglement since ER=EPR provides a concrete model of how QG might affect such problems, knowing however that such conclusions will always be conditional on the truth of ER=EPR. Our goal is to suggest a first avenue in which QG might inform the discussions on the compatibility between Humeanism and entanglement.\\
The paper's structure is the following: we give a simple presentation of the conflict between Humeanism and entanglement (\S\ref{2}). In particular, we reformulate this tension in a way that makes its subsequent connection with QG particularly intuitive, from a conceptual point of view. We then introduce the so-called ER=EPR conjecture, that we later use to elucidate how we can reconcile entanglement and Humeanism (\S\ref{3}). We explain how ER=EPR renders Humeanism compatible with entanglement (\S\ref{4}) and then we argue for the generality of ER=EPR and defend a metaphysics based on it, notwithstanding its conditional character (\S\ref{5}). In (\S\ref{6}) we conclude.


\section{Humeanism and Entanglement }\label{2}
In general, it can be quite difficult to pin down what exactly is meant by Humeanism, and we certainly do not want to take stock on a precise definition of this thesis. Hence, in the present paper, we rely on a simple characterisation of the core aspects of Humeanism. Following \cite{Maudlin2007-MAUTMW} Humeanism is the conjunction of these two doctrines:
\begin{itemize}
\item[Doctrine 1] \textbf{Separability}: The complete physical state of the world is determined by (supervenes on) the intrinsic physical state of each spacetime point (or each pointlike object)\footnote{Note that here we use the word object as a sui generis term, without commitments to any particular metaphysics of objects. Furthermore, from now on, for ease of discussion, we will, for the most part, refer only to spacetime points, when talking of Humeanism. The reference to pointlike objects is left implicit.} and the spatio-temporal relations between those points.
\item[Doctrine 2 ] \textbf{Physical Statism}: All facts about the world, including modal and nomological facts, are determined by its total physical state.
\end{itemize}
Given these two doctrines, let us now see why entanglement appears to be in tension with Humeanism. In particular, we show that entanglement explicitly violates \textbf{Separability}. The violation of \textbf{Separability} is because entangled states are not factorisable, that is, given a certain quantum state $\ket{\psi}$ for a composite system $\psi$ of two subsystems $\alpha$ and $\beta$, we cannot express it as a product state of the states of the two subsystems, that is, as a state of the form $\ket{\psi} = \ket{\alpha} \otimes \ket{\beta}$. The importance of product states for composite systems is that, whatever predictions one can derive from a product state, they are fully determined by the (intrinsic) state of its subsystems (equivalently, there is a common cause in the past which screens off any correlation in the product state). We cannot factorise the state of a composite system $\psi$ since there are correlations between the two subsystems $\alpha$ and $\beta$ that cannot be understood simply by taking into account the (intrinsic) states of the subsystems. Equivalently, there is no common cause which screens off the correlations.\footnote{This is particularly evident from Bell's inequalities \citep{Mermin1985IsTM}} Hence, these correlations are irreducible, i.e. there are no intrinsic states (that is, in this quantum mechanical case, pure states) that one can assign to the subsystems $\alpha$ and $\beta$ that can reproduce the predictions that one derives from an entangled state for the composite system $\psi$. A concrete example of such a state is the (in)famous singlet state, a state of two fermions (i.e.\ spin half particles), whose spins are entangled. The state of such a composite system is $\ket{\psi} = \frac{1}{\sqrt{2}}(\ket{\uparrow \downarrow} - \ket{\downarrow \uparrow})$. Being in this state entails that there is a correlation between the spins of the two particles. In particular, that their spins are anticorrelated, i.e.\ one particle has spin-up, the other has spin-down and vice versa. Indeed, it is this feature that engenders the conflict with \textbf{Separability}.\footnote{Though, most notably, this fact did not impress the 20th-century arch-Humean David Lewis \citep[pag. x-xi]{Lewis1986-LEWPPV-4}} As \citet[p. 51]{Maudlin2007-MAUTMW} says,
\begin{quotation}
\noindent if the principle of \textbf{Separability} holds, then each electron, occupying a region disjoint from the other, would have its intrinsic spin state, and the spin state of the composite system would be determined by the states of the particles taken individually, together with the spatiotemporal relations between them. But, no pure state for a single particle yields the same predictions as the Singlet state.
\end{quotation}
Note that here we are following \cite{Maudlin2007-MAUTMW} in finding the tension between Humeanism and entanglement at the level of \textbf{Separability}. However, such a problem only emerges insofar as one keeps fixed the assumption of \textbf{Physical Statism}. One might then be tempted to consider \textbf{Physical Statism} the problematic assumption for the Humean and retain \textbf{Separability} instead.\footnote{This view is briefly considered and rejected in \cite[section 3]{Bhogal2017-BHOWTH}. Furthermore, dropping \textbf{Physical Statism} seems to us to go against the core tenet of Humean metaphysics, i.e. that there are no necessary connections between distinct entities.} This move, i.e.\ to negate that the state of the world determines modal and nomological facts, implies that the correlations between entangled systems are essentially nomological. Giving a proper Humean understanding of this fact is then the main goal of this type of approaches.\footnote{A possible example of a view of this kind (at least in spirit), developed in the context of Bohmian mechanics, could be the so-called \textit{nomological interpretations of the wavefunction} \citep{Sol2019, Drr2013}.} However, for the remainder of this paper, we will bracket this type of concerns and assume that \textbf{Physical Statism} has to be retained and that it is \textbf{Separability} which is problematic for the Humean \textit{vis a vis} entanglement. Let us now step back slightly and recast this conflict in a way more conducive to its connection with QG. \\ 
The correlations involved in entangled states are particularly robust, in the sense that they are not merely accidental correlations between two subsystems, but are codified in the laws of nature of QM. They thus have, insofar as we take those laws to be a reliable guide at least to nomological modality, an important modal, and thus counterfactual, robustness.\footnote{Here by robustness we only refer to the counterfactual stability of entanglement correlations. We are not referring to the sense of robustness developed by \cite{Redhead1987-REDINA} as a way to examine the stability of entanglement.} Indeed, such counterfactual robustness of entanglement correlations has convinced some philosophers that we should understand these correlations as causal dependencies \citep{Maudlin2002-MAUQNA}. For simplicity of exposition, we too refer to entanglement correlations with talk of causal-like dependencies. Note, however, that for the notions that we introduce, this is not strictly necessary. The robustness of the correlations is sufficient to define all the notions that we use. In particular, nothing in our arguments hinges on some of the features typical of stronger accounts of causation, be them either energy-momentum exchange among causally related systems or causally significant interventions being possible. Since however, talking in causal terms renders the discussion much more straightforward, we make use of this talk. Those who disagree that entanglement correlations can be understood as encoding causal dependencies are free to substitute causality with robust counterfactual correlations. Indeed, outside of issues of linguistic simplicity, we encourage the reader to think, for the remainder of the paper, of all talk of causality as being concerned only with counterfactually robust correlations, and not necessarily with any stronger notion of causation.\\
Let us also note that we speak, quite liberally, as if it makes sense to locate quantum systems at spacetime points. This issue is controversial,\footnote{See \cite{pittphilsci1051}.} but we bracket such an issue, again for the sake of linguistic simplicity. We thus speak as if the location of quantum systems at points makes sense while noting that everything that we say continues to make sense even if we only had at our disposal regions of spacetime. However, we should note that while the core of our arguments is not influenced by whether or not quantum systems can be located at spacetime points, one might be sceptical that Humeanism can still make sense in the absence of definite location at spacetime points.\footnote{Though see \citep{Butterfield2006-BUTAPA, butterfield2005pointillisme} for a version of Humeanism that does not rely on any notion of spacetime points.} Given, however, that it is not settled in the literature that both these situations are the case, we proceed by bracketing these problems, speaking as if localisation at spacetime points of quantum systems made sense. At the same time, we think that it is useful to note that any satisfactory form of Humeanism should also address the concerns just mentioned.\\
Let us introduce the notion that is at the heart of our reformulation of the tension between Humeanism and entanglement.
\begin{itemize}\label{causal}
\item[\textbf{(CS)}] \textbf{Causal Structure}: given a theory $T$, we say that a causal structure according to the theory $T$ is given by a set $\Omega$ of spacetime points/pointlike objects (with their intrinsic physical state) and a relation $R$ which determines if two objects/points of spacetime can or cannot be causally related.\footnote{Which, we remind the reader, for us only means that there are robust counterfactual correlations.} \end{itemize}
\noindent
 Given this definition of causal structure, we can understand a particular causal structure as being determined through \textbf{(CS)} by three things: a theory $T$, a set of objects $\Omega$ (with their intrinsic physical states) and a relation $R$. For example, we might take as our physical theory $T$ GR, with its set of objects $\Omega$ being the spacetime points, and identify the relation $R$ in \textbf{(CS)} with the relation $R_{LC}$ of \textit{being connectable by a causal curve}, obtaining between two points of spacetime $p$ and $q$.\footnote{Note that is not obvious how GR should be interpreted. Here, however, for the sake of exposition, we speak explicitly of spacetime points. Nonetheless, it should be possible to carry over this discussion for other approaches to GR's ontology, at least insofar as a notion of object is admitted in the ontology.} We understand the relation $R_{LC}$ as obtaining if and only if there is a causal curve between the two points of spacetime $p$ and $q$. By causal curve, we mean a timelike or null curve (that is a curve which is always within the light-cone, border included, or, equivalently, a curve representing the worldline of an object whose speed never exceeds the speed of light). We call this the \textit{causal structure of (relativistic) spacetime}.\footnote{In what follows we omit the relativistic qualification as we always talk about relativistic spacetimes.} Thus, causal curves describe the causal structure of spacetime, in the sense that they determine which points can be in causal contact. Note that with this relation one automatically encodes the locality properties of relativity theory, since it follows that there cannot be spacelike separated points which are in causal contact. \\
Of course, for a given theory $T$, there might be more than one causal structure, i.e.\ more than one relation of causal connectedness, at least insofar as these relations are mutually compatible. However, since separability is only concerned with spacetime relations, the only relation of causal connectedness relevant for \textbf{Separability} is $R_{LC}$. Equivalently, \textbf{Separability} only relies on the causal structure of spacetime. For this reason, the definition above of the causal structure of spacetime allows us to redefine \textbf{Separability} in the following way:
\begin{itemize}
\item[Doctrine 1] \textbf{Separability${}^\dagger$}: The complete physical state of the world is determined by (supervenes on) the causal structure of spacetime.
\end{itemize}
Ultimately this is just an innocent way to recast \textbf{Separability} in terms of causal structures. Indeed, this is evident when we consider that, given any two spacetime points $p$ and $q$ (the objects with which \textbf{Separability} is concerned), the causal structure of spacetime, i.e.\ \textbf{(CS)} with $R_{LC}$ as the relation of causal connectedness, encodes the causally relevant spatiotemporal relations of $p$ and $q$ and their intrinsic physical states. However, this is just the structure with which \textbf{Separability} is concerned since it only deals with spacetime points, their intrinsic physical states, and their spacetime relations. Thus \textbf{Separability} and \textbf{Separability${}^\dagger$} are equivalent and, as consequence, we speak just of \textbf{Separability} henceforth.\footnote{Note that this equivalence holds also in the case of Galilean spacetime, assuming we use as the relation of causal connectedness the relation $R_G$ of \textit{being in spatial contact with}.}\\
However, an advantage of our way of recasting the problem, is that our definition of causal structure allows us to define a causal structure for entanglement. The relevant theory $T$ in the definition of \textbf{(CS)} would be QM, and the relation $R$ is identified with the relation $R_E$ of \textit{being entangled with}, connecting two quantum systems $\alpha$ and $\beta$. Indeed, entanglement determines robust correlations among systems,\footnote{As shown in \citep{Maudlin2002-MAUQNA, healey2016pragmatist}} and, in this paper, robust correlations are all that we mean when we say that there is a causal relation. Thus, since entanglement fixes a set of (cor)related objects which are in causal contact according to our framework, we can use it to construct a structure which fits into our definition \textbf{(CS)} of causal structure, by using entanglement as its generating relation $R_E$.\footnote{Note that, for the purposes of this paper, the causal structure of entanglement does not need to capture, in any interesting way, the overall causal structure of the world. This fact would imply the much stronger condition of entanglement fundamentalism \citep{pittphilsci17099}. We only need it to capture the structure of entanglement relations among quantum systems, as encoded in the robust counterfactual connection holding between them.} We can immediately see the sense in which in this framework we have a violation of \textbf{Separability}: one can have spacelike separated spacetime points (or regions, depending on our understanding of location for quantum systems) at which there are objects in entangled states. Thus, there are points causally separated for the causal structure of spacetime (defined by $R_{LC}$) but causally related for the causal structure defined by the entanglement relations $R_E$.\footnote{This only holds for entanglement between spacelike separated systems. From the present perspective, the case of timelike entanglement \citep{Olson:2011bq} does not create problems for the Humean since the two systems can be connected by a causal curve, explaining their correlation. For the remainder of the paper, whenever we speak of entanglement, we mean spacelike entanglement.} Note that this fact does not imply action-at-a-distance, since entanglement, and thus the relation $R_E$, only imply the existence of robust counterfactual connections which, however, do not necessitate any form of action at-a-distance.\footnote{As argued, for example, in \cite{myrvold2015lessons}. Indeed action-at-a-distance, as \cite[p. 292]{wallace2012emergent} notes, would be incompatible with relativity. However, in this article, we are interested in the incompatibility between Humeanism and non-separability, which we understand with the formalism of causal structures.} The violation of separability is then understood as the mismatch between the two causal structures since this mismatch entails that the complete physical state of the world cannot be determined by the causal structure of spacetime alone, contra \textbf{Separability}.\footnote{Observe that our discussion has been cast in the context of relativistic spacetime, i.e.\ the relation of causal connectedness in \textbf{(CS)} is given by $R_{LC}$. While this assumption is useful in the context of our paper, it is by no means required by the incompatibility between entanglement and Humeanism. Indeed one could recast the arguments of this section in the context of Galilean spacetime, by using as the relation of causal connectedness the relation $R_G$ of \textit{being in spatial contact with}, and the arguments would run parallel to those of this section.} Resolving this mismatch is the goal of the rest of this paper. To start, however, we need to introduce the specific formal and physical structures of the ER=EPR conjecture, which is the content of the following session.\footnote{Observe that this mismatch depends crucially on the fact that Humeanism relies only on the causal structure of spacetime, a fundamental assumption of the Humean view which we do not discuss in this paper.}


\section{Physical Framework}\label{3}
\cite{Maldacena:2013xja} point out the existence of a relation between entanglement and spacetime geometry, which they summed up with the slogan ER=EPR. ER refers to \cite{PhysRev.48.73}, a work regarding a novel solution of Einstein gravity which has been called Einstein-Rosen bridge (ERB henceforth). EPR refers to \cite{PhysRev.47.777}, the paper of the famous EPR paradox concerning quantum entanglement. In both papers, despite their different subject matter, locality appears to be challenged: ERB (wormholes) are non-local connections between black holes in GR; entanglement is a non-local correlation between particles in QM. \cite{Maldacena:2013xja} conjecture that these two phenomena are the same. The natural place to look for such connections between GR and QM is the physics of black holes because they are one of the few known objects displaying quantum gravitational effects since around the singularity inside the black hole GR breaks down and QG is needed to understand physics. In particular, we are interested in entanglement-like correlations between black holes. Does it make sense for a system of two black holes to be entangled? If this is the case, can entanglement of black holes be compared with ERBs?  

\subsection{Entanglement}\label{3.1}
\begin{figure}[h!]
\centering
\includegraphics[scale=0.6]{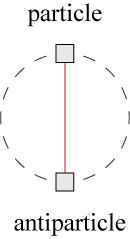} 
\caption{In the figure, the dotted line represents the propagation of a virtual particle or antiparticle, the boxes represent the particles/antiparticles and the red line represents the entanglement between them. We use these conventions hereafter.}
\label{fig:vacuum_bubble}
\end{figure}

\begin{figure}[h]
\centering
\includegraphics[scale=0.7]{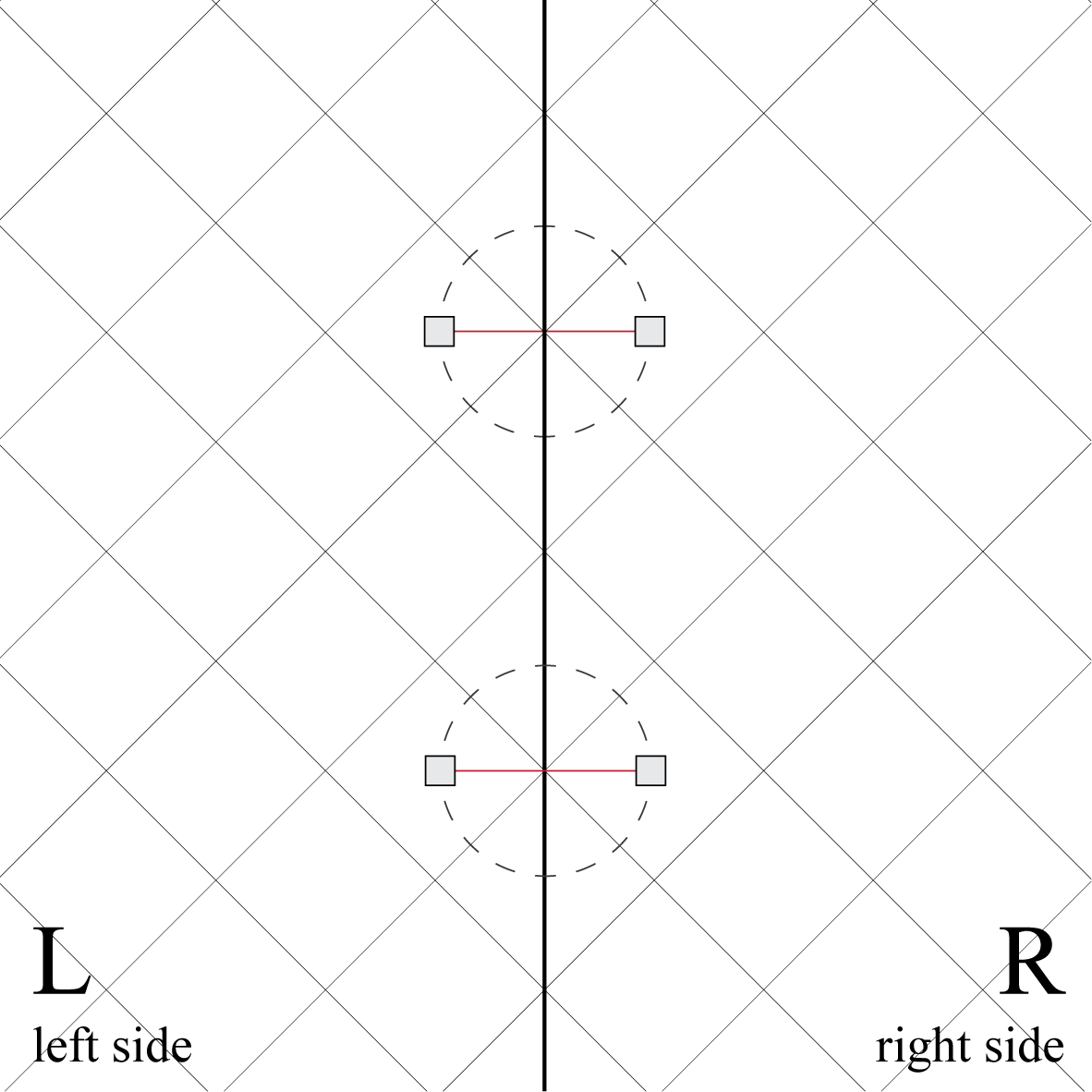} 
\caption{This figure describes spacetime equipped with a rhomboidal grid and divided into two halves. Some cells have some vacuum polarisation, others not.}
\label{fig:vacuum_entanglement}
\end{figure}

Let us define quantum entanglement so: given two arbitrary systems, a composite state $\ket\psi$ is said to be entangled if it is not a product state of the two subsystems. As we already discussed in \S\ref{2}, the simplest entangled system is a composite system of two maximally entangled qubits, i.e.\ a Bell pair, also known as \textit{singlet state}:
\beq
\ket \psi =\frac{1}{\sqrt 2}\left({\ket \uparrow}_1{ \ket \downarrow}_2 -{ \ket \downarrow }_1{\ket \uparrow}_2\right)~,
\eeq
where $1$ stands for the first qubit and $2$ stands for the second. A more relevant example for our purposes of entanglement comes from QFT. A key feature of this theory is that the vacuum is not simply empty space, but contains virtual particle-antiparticle pairs (see figure \ref{fig:vacuum_bubble}). These \textit{vacuum bubbles} immediately annihilate and are not measurable. 
 \\
The existence of such \textit{vacuum polarisation} straightforwardly leads to entanglement. Consider empty space and divide it into two halves: the left side $L$ and the right side $R$ (see figure \ref{fig:vacuum_entanglement}). Now take some cells, i.e.\ small regions of spacetime, near the boundary of the two halves. Consider then vacuum fluctuations across the boundary: if there is no particle on the left side then there is no particle on the right side; if there is a particle on the left side then there is its antiparticle on the right side. \\
 \\
Let us call $\ket 0$ the state where there is no particle and $\ket 1$ the state where a particle is present. The state which represents the system in each cell is the entangled state
\beq
\ket \psi = \frac{1}{\sqrt 2} \biggl({\ket 0}_L {\ket 0}_R + \ket{1}_L \ket{1}_R\biggr)~,
\eeq
where $L$ and $R$ refer to the left and the right side. By conformal invariance,\footnote{Conformal invariance means that the metric at a given spacetime point $x$ is invariant under local rescaling $\bar g_{\mu\nu} (x)= e^{2\omega (x) }g_{\mu\nu}(x)$. } a property of Minkowski spacetime, one can consider bigger and bigger cells, moving farther and farther away from the boundary. Thus, it is possible to find entanglement-like correlations between regions on the left and right sides even far away from the boundary. We can call these correlations \textit{ground state} or \textit{vacuum entanglement} \citep{Susskind:2016jjb}. In particular, if we take a quantum field to be a scalar field representing a coordinate system on spacetime, then it seems reasonable to say that vacuum entanglement captures a property of spacetime.

\subsection{Entangled Black Holes}\label{3.2}
\begin{figure}[h!]
\centering
\includegraphics[scale=1]{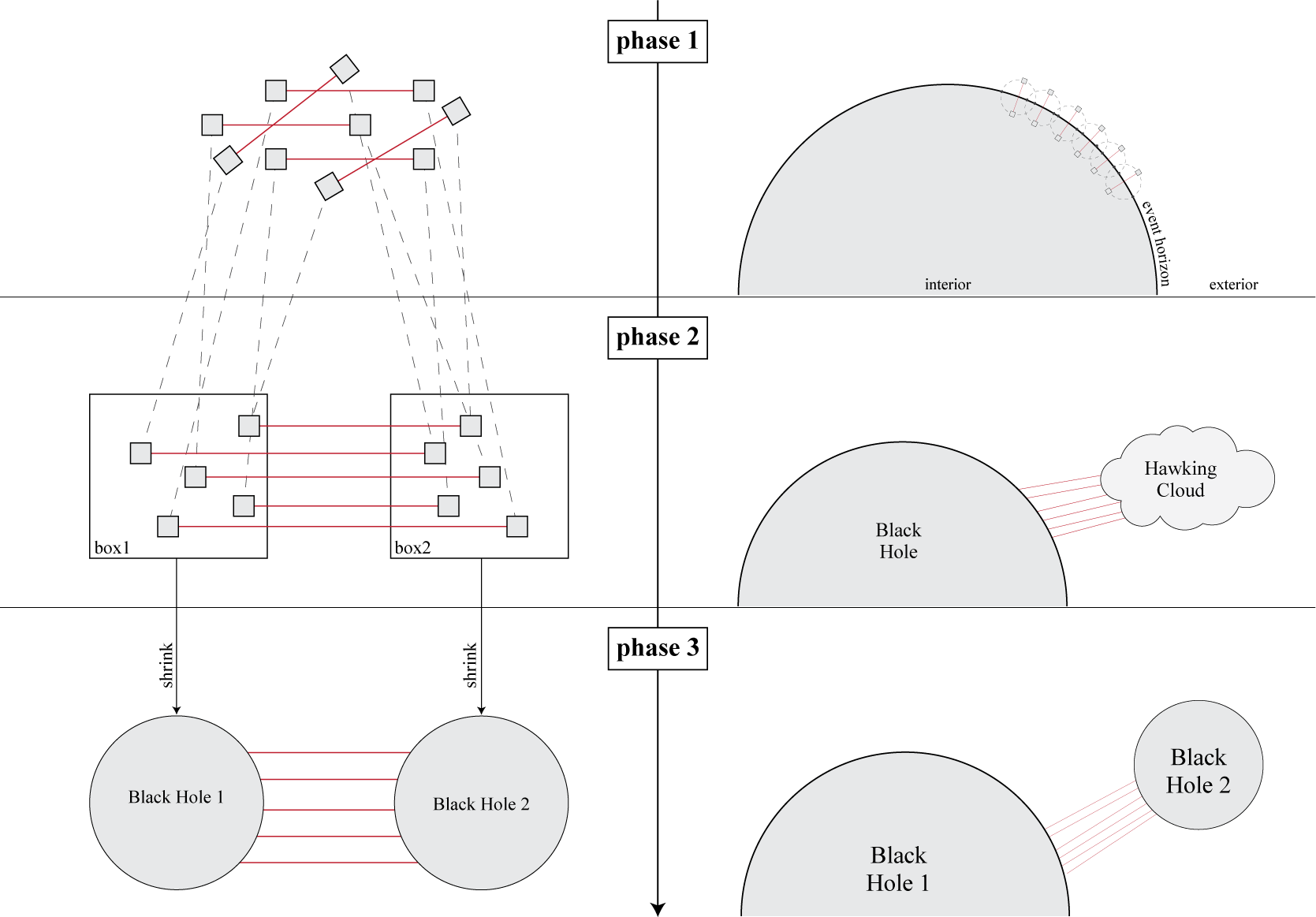} 
\caption{The figure represents two different ways to construct a pair of two entangled black holes. On the left, one starting from many Bell pairs, divide them into two boxes and shrinking them into two entangled black holes; on the right, we start with a black hole, collect its Hawking radiation and shrink the Hawking radiation into a second black hole.}
\label{fig:entangledBH2}
\end{figure}

So far we explained what we mean by entanglement, and how two entangled systems can be constructed in any quantum theory using either quantum entanglement or vacuum entanglement. Now we can turn to how it is possible to build up systems made of two entangled black holes.\footnote{Note that to do this we rely on the most natural assumption that black holes can be treated as quantum mechanical systems and thus can be entangled.} There are two ways of entangling black holes, using either quantum entanglement or vacuum entanglement (see figure \ref{fig:entangledBH2}). 
 \\
Take, for instance, quantum entanglement. Start by taking two boxes and a Bell pair, putting one of the two particles of the Bell pair into one box and the second particle into the second box. Then repeat this procedure for several Bell pairs. The result is two large scale objects, one in each box, entangled with each other. Then, one can, \textit{in principle}, create two entangled black holes just by shrinking the two boxes containing the entangled objects.\footnote{Of course this is merely a thought experiment. Indeed we do not take responsibility for the reader's attempts at constructing black holes and their consequences.} \\
The second way of constructing entangled black holes goes along the following lines: consider the event horizon of a black hole, and consider the empty space entanglement across the horizon, i.e.\ the entanglement created by dividing empty space into two halves as in figure \ref{fig:vacuum_entanglement}. The left part of figure \ref{fig:vacuum_entanglement} could be considered the black hole's interior, while the right part the exterior. The presence of a (virtual) particle inside the black hole would imply the presence of its antiparticle outside the horizon. The particle, being outside the horizon, is free to move away from the black hole and indeed, it does so for modes created at sufficiently high energy. This process is equivalent to the emission of a particle from the black hole. Since this process can happen for any number of particles, we end up with an ensemble of Hawking quanta, i.e.\ particles emitted through vacuum fluctuations across the horizon. This ensemble is called Hawking radiation. By construction then, the cloud of Hawking radiation is entangled with some regions of spacetime inside the black hole. Then one might take the cloud of Hawking radiation and shrink it, to form a second black hole. Therefore we end up with two entangled black holes. \\
These are two ways to realise (theoretically) a system of two entangled black holes. In the next section, we will look at non-local connectivity between black holes within the framework of GR.

\subsection{Einstein-Rosen Bridges \footnote{This subsection is rather mathematical. For background, the reader might want to consult \cite{Wald:1984rg} or \cite{Malament2012-MALTIT}. The motivation of this section is the lack of an appropriate mathematical introduction to wormholes in the philosophical literature. Those who are not interested in the mathematical details of wormholes might want to skip directly to the conclusion of the section. Keep in mind that the main takeaway of this section is that there are solutions of the vacuum Einstein equations which describe a situation in which two disconnected regions are connected through a geometrical bridge (the wormhole).}} \label{3.3}
\begin{figure}[h!]
\centering
\includegraphics[scale=0.8]{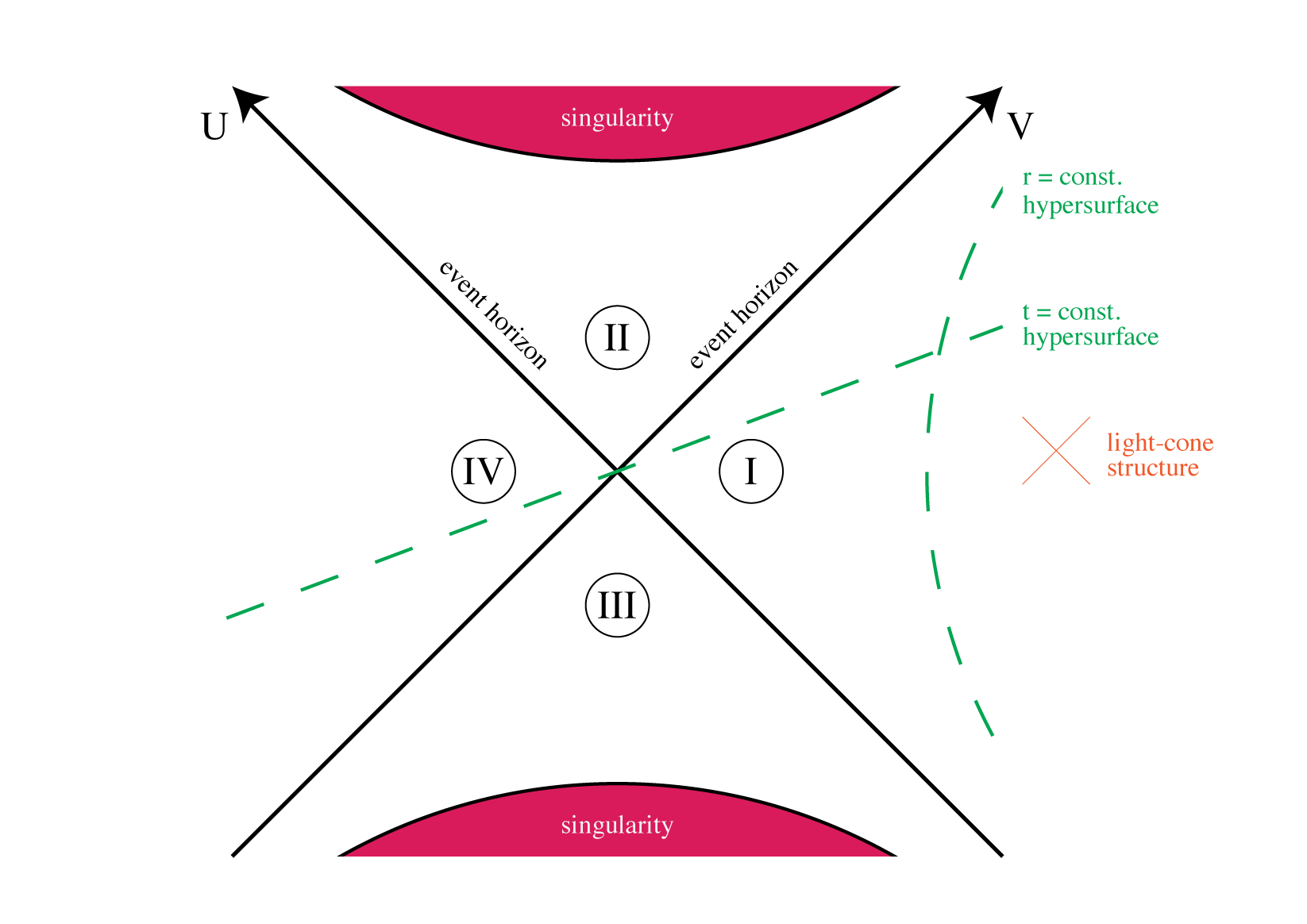} 
\caption{This is the Penrose diagram of the Kruskal solution.}
\label{fig:Kruscal_Penrose}
\end{figure}

\begin{figure}[h!]
\centering
\includegraphics[scale=0.8]{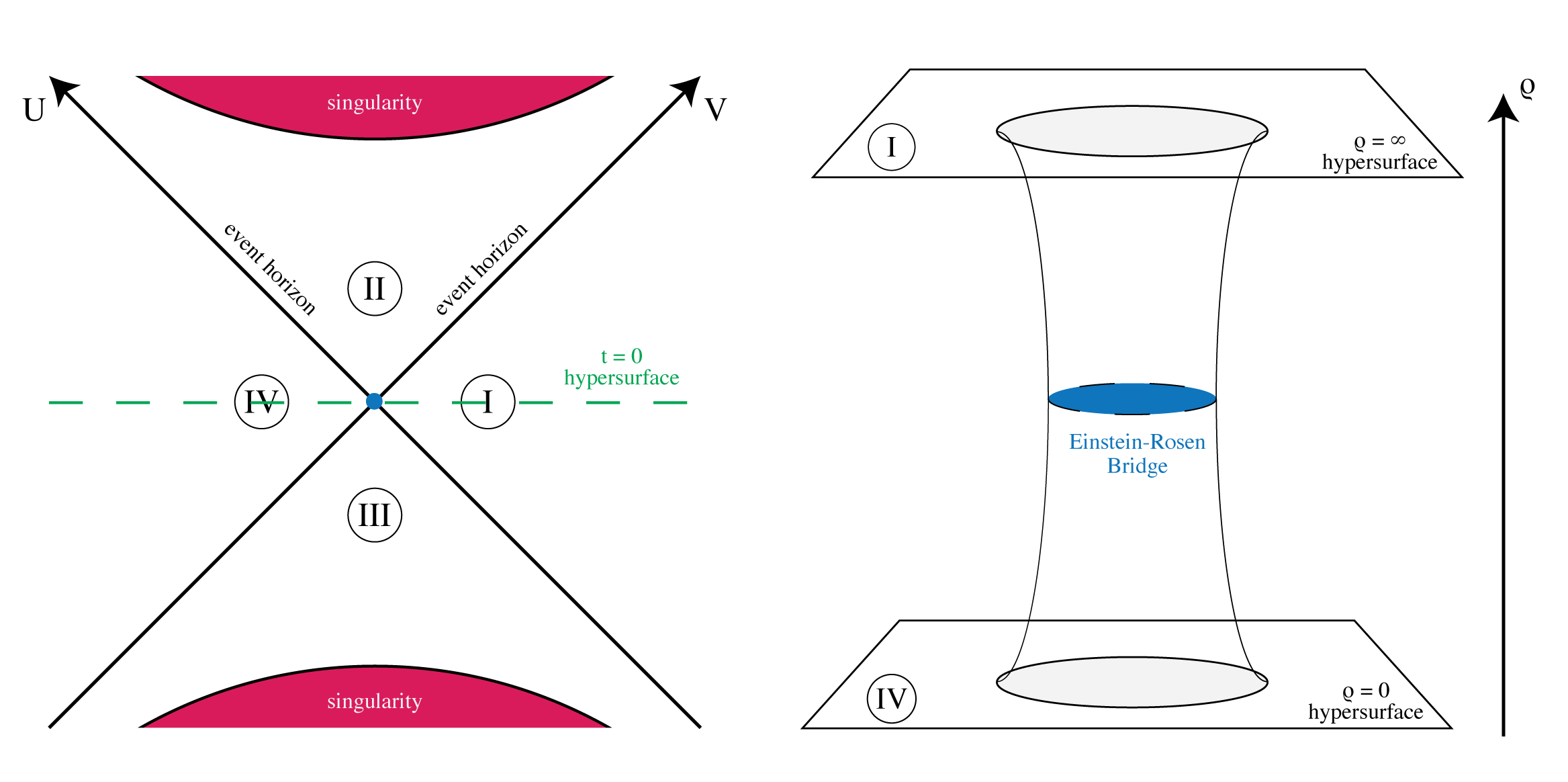} 
\caption{The left picture shows the Penrose diagram of the Kruskal extension. The right figure shows the $t=0$ hypersurface in the parametrisation with $\rho$. Region I and Region IV are respectively located at $\rho=\infty$ and $\rho=0$. These are two separate regions of space (it is evident that they are spacelike from the Penrose diagram), but a throat connects them. The blue dot in the Penrose diagram is a sphere $S^2$, which is the Einstein-Rosen bridge.}
\label{fig:ERbridge}
\end{figure}

The simplest black hole solution one can consider within GR's framework is the Schwarzschild solution, which describes a black hole without charge and angular momentum, and with spherical symmetry. A straightforward generalisation of the Schwarzschild solution is the Kruskal extension, which is derived just by performing a coordinate transformation on the Schwarzschild black hole metric. The geometrical structure described by the new set of coordinates is still a solution of the vacuum Einstein equations since the metric components are real analytic functions of the coordinates. Defining $r_\star = r+2M \log \left| \frac{r}{2M}-1 \right|$ (where $M$ is the mass of the black hole and $r$ is the radial coordinate of the Schwarzschild metric), the \textit{Kruskal-Szekeres} coordinates can be written down as follows
\beq\label{eq:KScoordinates}
U &=&-e^{-\frac{t-r_\star}{4M}}\\
V &=&e^{\frac{t+r_\star}{4M}}\\
\theta &=& \theta\\
\phi &=& \phi~,
\eeq
where, by construction, $U<0$ and $V>0$. The black hole metric in the Kruskal-Szekeres coordinate system is then
\beq\label{eq:kruscalmetric}
ds^2 = -\frac{32 M^3 e^{-r(U,V)/2M}}{r(U,V)}dU\,dV + r(U,V)^2d\Omega^2
\eeq 
where $r(U, V)$ is a solution of the equation 
\beq
UV=-e^{-r/(2M)}\left(\frac{r}{2M}-1\right)
\eeq
The function $r(U,V)$ can have definite values also for $U\geq0$ and for $V\leq0$, i.e.\ outside the domain in which $U$ and $V$ were defined in \eqref{eq:KScoordinates}. Since the metric components of \eqref{eq:kruscalmetric} depend only on $r(U, V)$, the Schwarzschild solution written in terms of Kruskal-Szekeres coordinates \eqref{eq:KScoordinates} can be analytically extended through the surfaces $U=0$ and $V=0$ to new regions of spacetime with $U>0$ and $V<0$. The new geometrical structure one obtains through these procedures is still a solution of the vacuum Einstein equations, and it has some interesting features which we can describe through its Penrose diagram (figure \ref{fig:Kruscal_Penrose}). 
 \\
The Penrose diagram of the Kruskal extension of the Schwarzschild solution underlines some interesting features. The solution has two horizons since $r=2M$ (the horizon's position in the Schwarzschild solution) corresponds either to $U=0$ or $V=0$. The diagram has four regions: 
\begin{itemize}
\item region I is the region for $r>2M$ of the Schwarzschild black hole, i.e.\ its exterior; 
\item region II is the interior of the Schwarzschild black hole, i.e.\ the region for $0<r<2M$. Since the Schwarzschild coordinates break down at the horizon, the appropriate set of coordinates for region II is the Eddington-Finkelstein coordinates $r^\star = r+ 2GM\ln\left|\frac{r}{2GM}-1\right|$, where $r$ is the Schwarzschild radial coordinate;
\item region III is a region describing a white hole (which however is not relevant for our discussion); 
\item region IV is a new region introduced by the analytic continuation performed above, where $r>2M$.
\end{itemize}
In both region I and region IV, $r>2M$, therefore both regions are described by the Schwarzschild solutions, and in particular, they can be understood as the exterior of two Schwarzschild black holes (since $r>2M$). So it is like there are two Schwarzschild black holes. Region I and Region IV are disconnected since they are spacelike separated, so it is impossible for an observer in Region I to send a signal in Region IV. The green lines represent r-constant and t-constant hypersurfaces. Moreover, since $V/U=-e^{t/(2M)}$, a surface of constant $t$ in the new coordinates is just a straight line through the origin of the Penrose diagram, which extends naturally from Region I to Region IV. One can then ask: which is the geometry of these constant time slices? Another change of coordinates is needed to investigate their geometry. We define the new coordinate $\rho$ in the following way:
\beq
r=\rho+M+\frac{M^2}{4\rho}
\eeq
This equation has two solutions, one for $\rho>M/2$ (which describes Region I) and one for $0<\rho<M/2$ (which describes Region IV). The Schwarzschild metric, written in this new coordinates, and at constant time, is then
\beq\label{eq:wormhole}
ds^2 = \left(1+\frac{M}{2\rho}\right)^4\left(d\rho^2+\rho d\Omega^2\right)
\eeq
This geometry has two asymptotically flat regions for $\rho\rightarrow \infty$ and $\rho \rightarrow 0$ connected by a throat of minimum radius $2M$. So a surface of constant time in the Kruskal metric is an Einstein-Rosen bridge (figure \ref{fig:ERbridge}). 
 \\
A bridge geometrically connects these two disconnected regions of space, i.e.\ there is a continuous path that connects the two disconnected black holes' interior. Two completely disconnected worlds are connected by an ERB.\\
To conclude, we just showed a solution of the vacuum Einstein equation which describes a situation in which two disconnected regions are connected through a new geometrical structure, i.e.\ a wormhole. Since there is no causal curve connecting the two regions, the ERB is a form of non-local connectivity. The bridge has two fundamental properties: first of all, it is pure empty space, i.e.\ it is not a fancy region of spacetime with some fancy properties, it is made up only of geometry; moreover, it is not traversable, i.e.\ one cannot walk across the bridge since it would exceed the speed of light.

\subsection{ER=EPR}\label{3.4}
\begin{figure}[ht]
\centering
\includegraphics[scale=0.8]{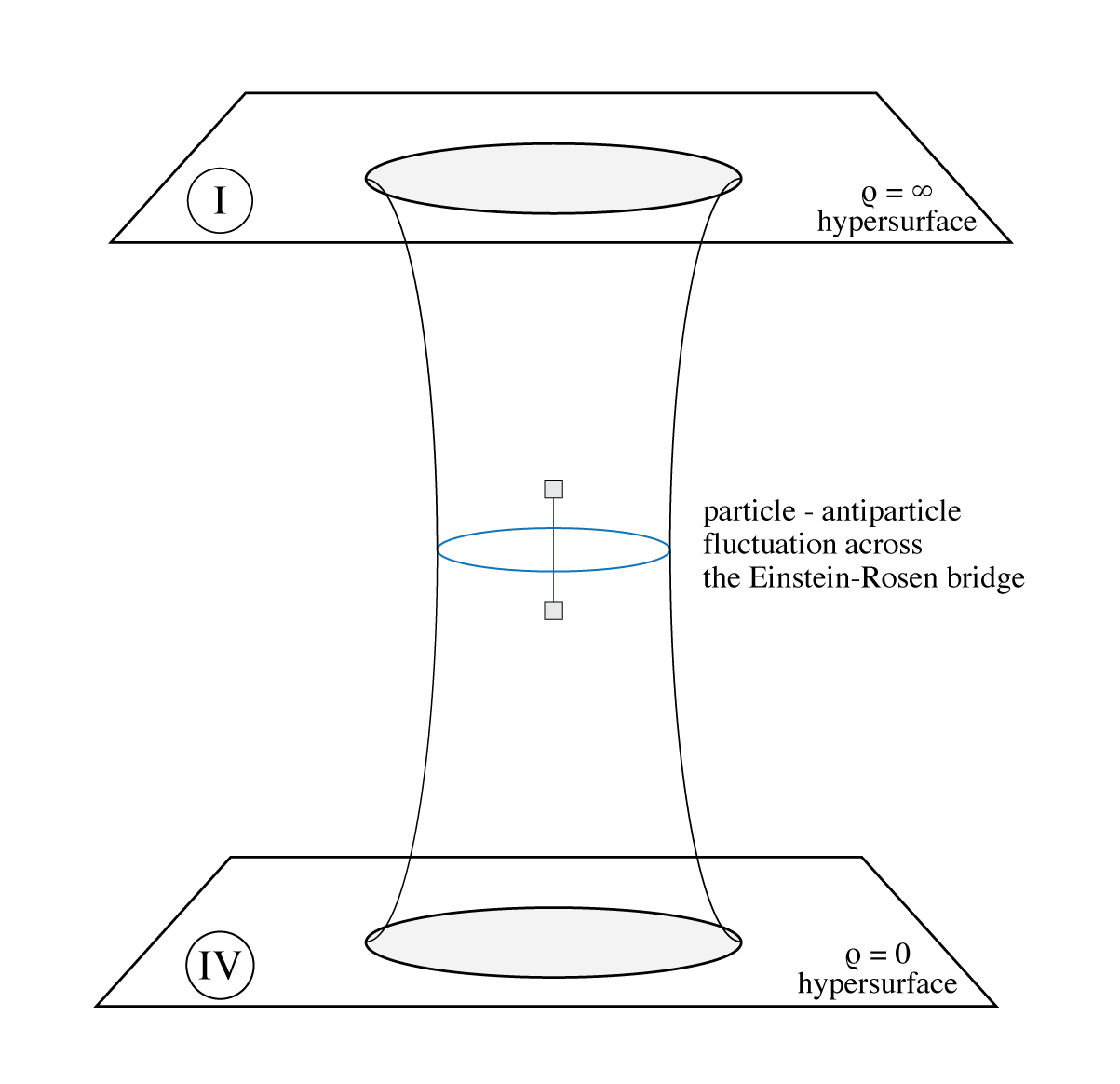} 
\caption{The figure shows empty-space entanglement across an ERB.}
\label{fig:EntERB}
\end{figure}

\begin{figure}[h!]
\centering
\includegraphics[scale=0.8]{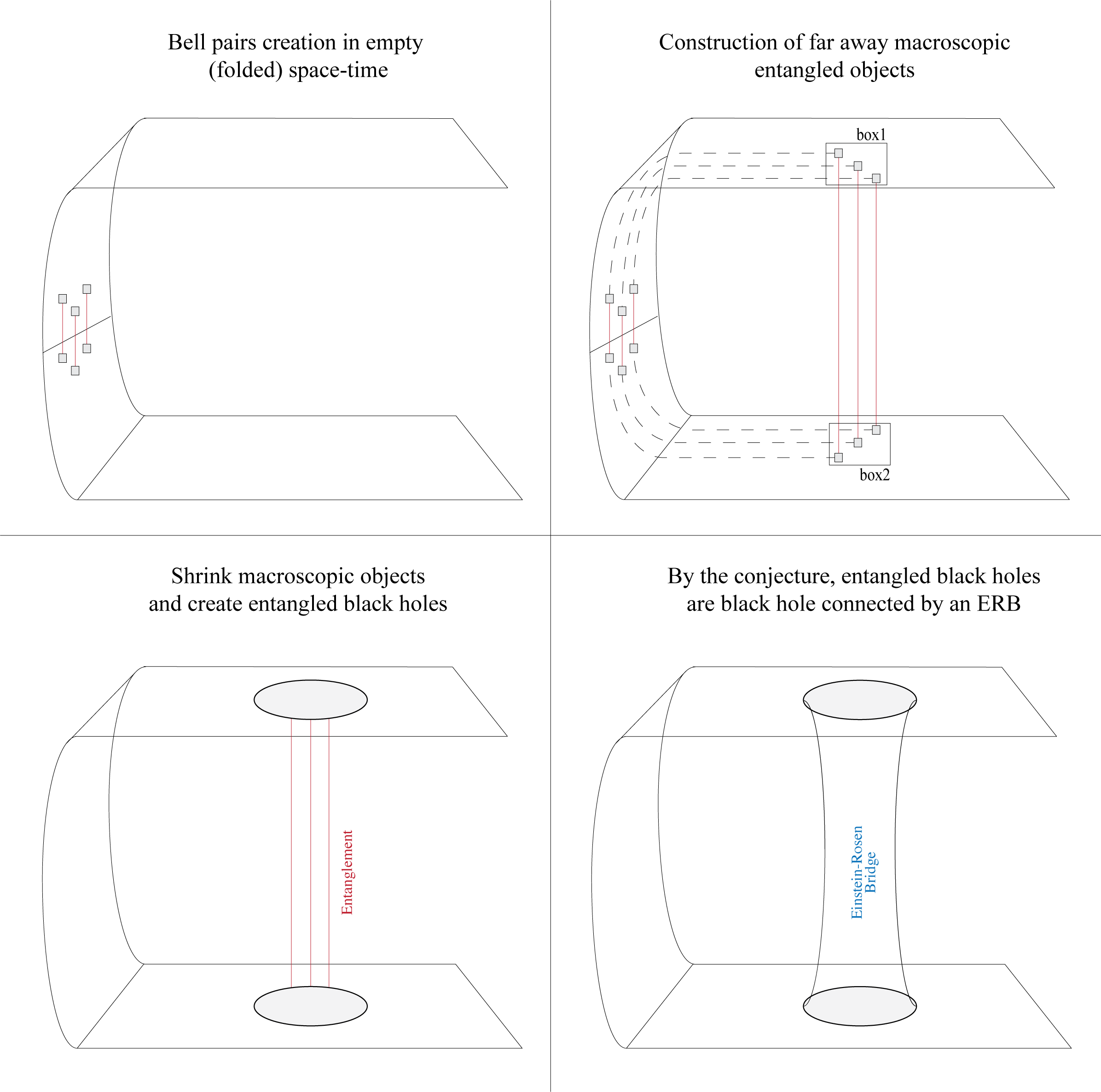} 
\caption{This figure gives a representation of how the modest version of ER=EPR can come about. One should read the panel as ordered from left to right, and from top to bottom.}
\label{fig:EREPR1}
\end{figure}

So far, we have seen that black holes can, in principle, be connected through two prima facie significantly different kinds of non-local connectivities: entanglement and ERBs. In this section, we investigate the relations between them. We learnt that ERBs are solutions of the vacuum Einstein equations, which means that they are purely geometrical structures. The fact that they are purely geometrical means the bridge itself is pure empty space. Since it is pure empty space, and since we showed that pure empty space in the quantum theory is intrinsically entangled (via vacuum entanglement), we can consider vacuum entanglement between the two sides of the bridge (figure \ref{fig:EntERB}).
 \\
This procedure suggests that two black holes that are connected through an Einstein-Rosen bridge are also entangled. This simple reasoning lead \cite{Maldacena:2013xja} to formulate what \cite{Susskind:2016jjb} calls the modest view of the ER=EPR conjecture: 
\begin{itemize}
\item[\textbf{(MV)}] \textbf{Modest View}: Black holes connected by an Einstein-Rosen bridge are entangled. Entangled black holes are connected by an Einstein-Rosen bridge.\footnote{Note that the previous argument establishes only the first clause of \textbf{(MV)}. The second clause is the most speculative one.}
\end{itemize}
That is to say that, when we consider entanglement between two spacelike macroscopic objects and those macroscopic objects get dense, then the entanglement turns into new regions of spacetime. Those regions would not have been there if the systems were not entangled. In other words, there is a one-to-one correspondence between wormholes and entangled black holes, i.e.\ any entanglement between two black holes corresponds to one particular wormhole geometry and vice versa. The diagram in figure \ref{fig:EREPR1} can pictorially describe the modest view of the ER=EPR conjecture.
 \\
We may ask the following question: is the equivalence between entanglement and ERBs restricted to the case of black holes? Or, in other words, given two entangled particles, does their entanglement turn into a new region of spacetime which connects the two particles? \cite{Susskind:2016jjb} calls this the ambitious view of the ER=EPR conjecture:
\begin{itemize}
\item[\textbf{(AV)}] \textbf{Ambitious View}: Some future conception of quantum geometry will allow us to think of two entangled systems as being connected by a Planckian wormhole. 
\end{itemize}
In this definition by \textit{Planckian} wormhole it is meant a wormhole of Planckian size, i.e.\ a wormhole for which the mass in \eqref{eq:wormhole} is of the same order of the Planck mass and thus invisible for standard experiments dealing with quantum entanglement. \textbf{(AV)} then states a complete equivalence between entanglement and ERBs. \textbf{(AV)} generalises \textbf{(MV)} by extending the correspondence between entangled black holes and wormhole geometry to any entangled systems, i.e.\ there is a one-to-one correspondence between entangled systems and wormhole geometry.

 \subsection{Fungibility of Entanglement}\label{3.5}
The ER=EPR conjecture allows us to think about entanglement in a very different way. \cite{Susskind:2016jjb} describes entanglement as a \textit{fungible resource}. 
\begin{quotation}
\noindent The idea can be stated in terms of entanglement being a "fungible resource." Entanglement is a resource because it is useful for carrying out certain communication tasks such as teleportation. It is fungible because like energy, which comes in different forms - electrical, mechanical, chemical, etc. - entanglement also comes in many forms which can be transformed into one another. Some forms of entanglement: ground state or vacuum entanglement, entangled particles, Einstein-Rosen bridges, [\dots]. What about the conservation of the resource? Energy is conserved, but entanglement is not, except under special circumstances. If two systems are distantly separated so that they can't interact, then the entanglement between them is conserved under independent local unitary transformations. Thus if Alice and Bob, who are far from one another, are each in control of two halves of an entangled system, the unitary manipulations they do on their own shares cannot change the entanglement entropy.
\end{quotation}
To understand this, let us recall a simple fact that we already mentioned before regarding entangled systems: how to create macroscopic objects displaying entanglement correlations (see figure \ref{fig:entangledBH2}). To create a macroscopic object displaying entanglement correlations, we need to take a pair of quantum systems in an entangled state and put each of them in a box. We can repeat this process and then, after adding a sufficiently large number of particles to the boxes, we end up with two entangled macroscopic objects, one for each box. We could then decide to shrink the size of the two boxes and, if we make them small enough, they become entangled black holes. However, since we are in the context of entangled black holes, \textbf{(MV)} applies. Therefore, we can equivalently regard the two black holes as being connected by a wormhole, whose interior geometry encodes their state's entanglement properties. In this way, we have seen how we can transform two entangled systems in two entangled black holes, thus susceptible to \textbf{(MV)}. In this sense, we can say that fungibility is the property which makes possible the transformation from vacuum entanglement to quantum entanglement, to ERBs.


\section{Humeanism and ER=EPR}\label{4}
As we showed in \S\ref{2}, the violation of \textbf{Separability}, i.e.\ that entanglement relations cannot be understood in terms of the intrinsic states of spacetime points/pointlike objects, together with their spacetime relations, is the cause of the conflict between entanglement and Humeanism. Such a violation is understood as entanglement and spacetime inducing two incompatible causal structures. These two causal structures are different because there are events causally connected in one which are causally disconnected in the other. A straightforward strategy to solve the problem might be eliminating entanglement relations or reducing them to some intrinsic property of quantum objects. Since entanglement is one of the fundamental discoveries of QM, and since we assume a form of entanglement realism,\footnote{Which, we remind the reader, we understand roughly speaking as the view that there are relations constitutive of entanglement in our fundamental ontology, i.e.\ we think that entanglement relations are part of the metaphysical structure of the world.} we think a better strategy is to search for a physical theory that resolves the inconsistency between causal structures induced by spacetime and entanglement. This section aims to argue that ER=EPR gives us the resources to solve the mismatch between these two causal structures. We thus submit that ER=EPR might provide a possible framework to overcome the conflict between entanglement and Humeanism in the context of QG. Or, at the very least, it provides a useful example of how one might think about these issues in QG.\\
ER=EPR harmonises the two causal structures by unifying a general relativistic spacetime with the entanglement of QM. The two structures are harmonised by associating to any two entangled quantum systems an ERB connecting them, thus assigning a spacetime connection to any entanglement correlation. The wormhole opens the possibility to create a path \textit{in spacetime} which connects the two entangled quantum systems. This new type of connectivity allows us to supplement the standard spacetime relations (light-cone structure, i.e.\ the relation $R_{LC}$ of \textit{being connectable by a causal curve}).\\
To formalise the view emerging from the ER=EPR conjecture, starting from our notion of causal structure \textbf{(CS)}, we generalise the notion of the causal structure of spacetime to incorporate this new kind of connectivity introduced by the ER=EPR conjecture. The new definition of the causal structure of spacetime is determined by the relation $R_{ERB}$ of \textit{being non trivially connectable in spacetime}. We say that there is a non-trivial connection in spacetime if there is (i) \textit{a continuous path in spacetime} connecting two points $p$ and $q$ such that (ii) the presence of this path determines whether or not there is a \textit{robust counterfactual correlation} between the systems/points it connects. We call this new causal structure \textit{the generalised causal structure of spacetime}.\\
Note that, if we go back to the context of standard Minkowski spacetime, the generalised causal structure of spacetime (and the notion of \textbf{Separability} defined by $R_{ERB}$) are equivalent to the notions of the causal structure of spacetime and \textbf{Separability} defined in terms of $R_{LC}$, i.e.\ causal curves. Indeed, in Minkowski spacetime, the only paths which can determine whether or not there are robust counterfactual correlations are causal curves, and thus we recover our original definition (and with it the problem of violations of \textbf{Separability} by entanglement). The real advantage of using the generalised causal structure of spacetime is that it captures the causal structure of the spacetimes emerging from ER=EPR. Since the ERBs determine whether two systems are entangled, and in which entangled state they are, they determine whether or not there is a robust counterfactual correlation between the two systems. Thus we can consider the ERBs as non-trivial paths in spacetime, despite being spacelike. To better understand the argument and the nature of this new kind of connectivity encoded by wormholes, let us now consider the ambitious version \textbf{(AV)} and the modest versions \textbf{(MV)} of ER=EPR individually. 

\subsection{Ambitious ER=EPR}\label{4.1}
That the two causal structures are harmonised is particularly evident in \textbf{(AV)}, since for any two entangled systems we have an associated wormhole connecting them. It is, however, essential to be clear on what is going on in this case. What \textbf{(AV)} says is that, given two entangled quantum systems, they are equivalent to a wormhole. What is important to note, then, is that we are not claiming that there is a wormhole connecting the two systems, but rather that the two entangled systems are equivalent to the wormhole: entanglement is equivalent to a specific type of spacetime connection, encoded via an ERB. The mismatch between the causal structure of entanglement and spacetime is resolved by this equivalence, which allows us to represent entanglement relations with non-trivial spacetime connections, i.e.\ ERBs, as captured by the generalised causal structure of spacetime. Thus Humeanism is again viable since specifying all the non-trivial spacetime relations in \textbf{(AV)} is equivalent to specifying the entanglement relations. We can determine the state of the world from the generalised causal structure of spacetime, satisfying \textbf{Separability}.\\
Indeed our reformulation of the problem in terms of causal structures becomes particularly useful in allowing us to see how to resolve it. What \textbf{(AV)} allows us to do is to construct a new causal structure, i.e.\ the generalised causal structure of spacetime defined by the relation $R_{ERB}$ of \textit{being non trivially connected in spacetime}. In the generalised causal structure of spacetime, both the causal structure defined in terms of $R_{LC}$, i.e.\ causal curves, and the causal structure defined by $R_E$, i.e.\ entanglement relations, can be consistently embedded. Furthermore, this causal structure is still ultimately a \textit{spacetime} causal structure, given that the relation $R_{ERB}$ is still a spacetime relation. Let us briefly expand on this point. As noted in \cite{brown2005physical}, understanding what exactly counts as a piece of spacetime structure is far from obvious. An influential account of how to identify spacetime structure, starting from the work of Brown himself, is that of \cite{knox2013effective}, who uses it to articulate her preferred version of spacetime functionalism.\footnote{For discussion on spacetime functionalism see \citet{knox2013effective, read2019limitations,baker2020knox,butterfield2020functionalism, lam2020spacetime}.} In particular, \citet[p. 122]{knox2019physical} characterises her criterion by saying that \enquote{[\dots] spacetime is whatever serves to define a structure of inertial frames, where inertial frames are those in whose coordinates the laws governing interactions[footnote suppressed] take a simple form (that is universal insofar as curvature may be ignored), and with respect to which free bodies move with constant velocity.}\footnote{\cite{baker2020knox} suggests that our concept of spacetime is a \textit{cluster concept}, and that \cite{knox2013effective}'s criterion is only one of the possible alternative criteria for what counts as spacetime structure.} To see why $R_{ERB}$ counts as a piece of spacetime structure from this point of view, it is simpler to consider the firewall paradox \citep{Almheiri_2013}. In that context, \cite{Almheiri_2013} constructed a situation in which the equivalence principle of GR is violated, and thus no local inertial frame in the sense of \cite{knox2013effective} can be defined. ER=EPR, and thus $R_{ERB}$, were introduced for the explicit purpose of avoiding this situation and ensuring the validity of the equivalence principle, a task accomplished with remarkable success \citep{Maldacena:2013xja, penington2019entanglement}. Thus, it is safe to conclude that $R_{ERB}$ plays a crucial role in determining the structure of local inertial frames and that, following \cite{knox2013effective}'s criterion, it counts as a piece of spacetime structure. \\
The fundamental new tools that \textbf{(AV)} puts at our disposal to reconcile the structure of spacetime with that of entanglement are (i) for each entangled state there is an ERB, and (ii) the features of the entangled state are encoded in the geometry of the wormhole. Thus, in the context of \textbf{(AV)}, Humeanism is a viable metaphysical picture of the world again.

\subsection{Modest ER=EPR}\label{4.2}
We can now turn to \textbf{(MV)} and see how in this case entanglement and spacetime can be harmonised, and thus Humeanism saved. Here, the situation is a bit more complicated than in the case of \textbf{(AV)}, since we cannot directly embed the two causal structures into a larger one, not in general at least. Indeed, in \textbf{(MV)}, we only have a correspondence between entanglement and wormholes for the case of black holes, not for any two quantum systems. This limited correspondence does not allow us to resolve the conflict between entanglement and spacetime as quickly but requires us to be more careful in treating the two causal structures. The fundamental concept that allows us to resolve the problems of Humeanism also in \textbf{(MV)} is that of the fungibility of entanglement explained in section \S\ref{3.5}. In this context, in particular, what we mean when we say that entanglement is a fungible resource is that we can transform entangled systems in entangled black holes which are then, by \textbf{(MV)}, connected by wormholes, with the same properties of the original quantum systems. This connection is the core of what we mean, in \textbf{(MV)}, when we say that entanglement is fungible. \\
While for \textbf{(AV)}, as noted above, we can simply embed the two causal structures, of spacetime and entanglement, in the generalised causal structure of spacetime, this is not so for \textbf{(MV)}, since entanglement and wormholes are not identified. However, the fungibility of entanglement shows that, while not identical, the two causal structures are compatible in a robust sense. Indeed, while not every entanglement relation corresponds to an ERB, we can transform with a sequence of unitary transformation two entangled systems into two black holes connected by a wormhole, with the same properties of the entangled pair, provided that we have a sufficient amount of entangled pairs to carry out constructions along the lines of \S\ref{3.2}.\footnote{For example, what we see here would not be the case in a world where there is a single pair of entangled particles.} For any world with, for example, two entangled systems, there is an equivalent\footnote{Note that here, by equivalent, we mean that there is a unitary transformation which connects world 1 with world 2. Therefore there is no information available in 1 that is not available in 2, and vice versa. For a defence of unitary equivalence as a sufficient criterium for physical equivalence in quantum theory see \citep{Baker:2015gmy,halvorson2007algebraic}.} world with the same overall physical state. There, however, entanglement has been substituted by a wormhole with the same properties. Thus, we can see that even in \textbf{(MV)}, Humeanism and entanglement can be reconciled. Fixing spacetime relations, even in the more general sense of $R_{ERB}$, i.e.\ non-trivial spacetime connections, might not be enough to determine the physical state of the world since entanglement still does not fall within the purview of this causal structure.
Nonetheless, by a finite number of unitary transformations, what we can do is change this entangled system into an equivalent one in which the generalised causal structure of spacetime is sufficient to determine the state of the world, without ever changing it.\\
The version of Humeanism that emerges is a somewhat weaker one, but still adequate in our opinion to the spirit of Humeanism. While it might not be true that the intrinsic state of spacetime points and their spacetime relations (equivalently, the generalised causal structure of spacetime) is sufficient to determine the physical state of the world, it is nevertheless the case that there is an equivalent world whose generalised spacetime causal structure does determine the global state of the world.\footnote{Again, provided that we have enough entangled pairs.} By equivalent, we mean that we can construct the generalised causal structure via a finite sequence of unitary transformations on the two (apparently) contrasting causal structures.\\
In fewer words, we can take an entangled state and act on it to transform into a wormhole (with the same properties). Thus, there cannot be any deep conflict between spacetime and entanglement's causal structures, only an apparent one. The dictionary between these two situations, provided by the fungibility of entanglement, guarantees that they are compatible, and thus that entanglement cannot threaten \textbf{Separability} in any significant way.\footnote{For otherwise, also wormholes would, but this would not make sense since the relations that they encode are ultimately spacetime relations.} Thus Humeanism, albeit somewhat weakened, can be saved even if \textbf{(MV)} is retained.


\section{In Defence of ER=EPR}\label{5}
In this section, we defend the generality of ER=EPR and the viability of constructing a metaphysics based on it. First of all, we observe that doing metaphysics on speculative theories of QG is by now a common practice in the literature (\S\ref{5.1}), even though the resulting metaphysics is equally speculative. The core of the section is the defence of the claim that the ER=EPR conjecture is rather general since it relies only on gravity's semiclassical features (\S\ref{5.2}). We then conclude by noting that wormholes being non-traversable is not a problem for our view (\S\ref{5.3}).

\subsection{ER=EPR and Metaphysics}\label{5.1}
The first, and most natural, criticism of our view is that it crucially depends on the ER=EPR conjecture, a particular, narrow and ultimately unproven conjecture regarding a possible future theory of QG. How can we draw metaphysical lessons from such a speculative and uncertain physics? First of all, let us briefly note that one might object that the dependence on a conjecture such as ER=EPR renders our view empirically falsifiable. In keeping with the methodology of naturalised metaphysics \citep{ladyman2007every}, empirical falsifiability should be one of the main features of an acceptable metaphysical position. From this point of view what seemed a bug of our position, should instead be seen as a feature. While we broadly agree with the perspective of naturalised metaphysics, however, we take that the criticism articulated above is still excessive. While we certainly agree that ER=EPR is ultimately still a conjecture, it is not nearly as far fetched as might appear at first sight. In particular, various experiments have been proposed in recent years which might test various aspects of the ER=EPR conjecture, which might soon be concretely performed in a lab \citep{nezami2021quantum,brown2021quantum, PhysRevD.102.066004}. Moreover, recent advances in quantum black holes have shown that various black hole models display the features characteristic of ER=EPR \citep{Papadodimas:2012aq, Almheiri_2019, penington2019entanglement, Almheiri_2020b}, some of which derived directly from the gravitational path integral \citep{penington2019replica, Almheiri_2020, almheiri2020entropy}. For these reasons, it seems to us that it is reasonable to begin the project of elucidating some of the metaphysical consequences that ER=EPR seems to suggest about the world, keeping in mind, of course, the unproven nature of the conjecture. Indeed, this type of metaphysical project should be understood as trying to articulate what the (still speculative) physics of QG might be ultimately telling us about reality. Something that has become, if not ordinary, at least broadly accepted in the literature on the metaphysical foundations of QG \citep{Matarese:2019bxo, vistarini2019emergence, pittphilsci15902, le2020string, pittphilsci17099}. Moreover, the ER=EPR conjecture, and thus the metaphysics based on it, does not rely for its formulation on the details of the (yet to be discovered) theory of QG. Instead, it relies only on (general features of) its expected semiclassical limit and GR's spacetime picture, as we argue in the next section.

\subsection{ER=EPR and Semiclassical Gravity}\label{5.2}
Let us now substantiate the claim that ER=EPR does not depend on the details of any specific theory of QG. Indeed, ER=EPR relies only on the semiclassical structure of spacetime. As such, insofar as the conjecture is correct, any QG theory should include it in its semiclassical description of reality.\footnote{To be more precise \textbf{(MV)}, being just a claim about the semiclassical structure of gravity, is independent of QG. On the other hand, known models instantiating the features of \textbf{(AV)} typically rely on the AdS/CFT correspondence. However, it seems that AdS/CFT can be realised in many different approaches to QG, from string theory \citep{Maldacena:1997re} to loop QG \citep{Han_2017}, and group field theory \citep{chirco2019group}.} \\
First of all, let us remember the context in which the ER=EPR conjecture emerged: as a(n ingredient of a) solution to the AMPS paradox \citep{Almheiri_2013}, also called the firewall paradox, the latest incarnation of the (in)famous information loss problem for black holes \citep{Wallace:2017wzs}. In the specific context of the eternal black hole, ER=EPR works beautifully in resolving the paradox, as shown in \cite{Maldacena:2013xja}. Furthermore, in such a model, constructed within the AdS/CFT correspondence \citep{Maldacena:1997re},\footnote{See \cite{Ammon:2015wua} for a useful review.} one can prove the conjecture to be true: entanglement between the two horizons (which, from the CFT perspective, are described by two entangled CFTs) of the eternal black hole is equivalent to a wormhole connecting them. ER=EPR, then, is not only a fundamental ingredient in some of the leading solutions to the firewall paradox,\footnote{For discussion of various approaches to the firewall paradox, and to the black hole information paradox more generally, see \cite{Harlow:2014yka}} but it has already shown to be valid in some specific models of QG, based on the AdS/CFT duality.\\
Indeed, more than this can be said. Recent advances in the study of the black hole information paradox have shown that the basic structure of the ER=EPR conjecture can be derived already form the path integral of semiclassical gravity \citep{penington2019replica, Almheiri_2020, almheiri2020entropy}.\footnote{More precisely, an explicit derivation has been provided for a two-dimensional theory of gravity, the JT gravity of \citet{TEITELBOIM198341, JACKIW1985343} (see \cite{saad2019jt} for non-perturbatively well-defined formulation). However, the construction is expected to generalise to more general cases, though see \cite{fletcher2018would} for subtleties on the connection between two-dimensional theories of gravity and higher dimensional ones.} This fact shows that ER=EPR is independent of the specific details of any full theory of QG, given that it is reasonable to expect that any successful theories of QG will recover the path integral of semiclassical gravity, in the appropriate limit. Indeed, this computation gives some evidence in favour of the idea that any theory of QG will include the realisation of the ER=EPR conjecture in its semiclassical limit. \\
Thus, the features on which we rely in our arguments can be expected to be a part of the semiclassical structure of a future theory of QG, lending a reasonable, though not absolute, degree of credibility to the ER=EPR conjecture and the metaphysics based upon it. Indeed, insofar as we are interested in ER=EPR as a possible guide as to how issues regarding Humeanism might be treated in the context of QG, it seems that the results that we have summarised in this section make the features central to the conjecture and to our argument robust enough that we can at least take them into account for metaphysical theorising. Ultimately, however, any results obtained in this way are conditional on the eventual confirmation of the ER=EPR conjecture or any possible cousin of it that might be instantiated in a future theory of QG.

\subsection{ER=EPR and Non-Traversable Wormholes}\label{5.3}
Let us turn to an objection to the metaphysical proposal that we are advancing. One might be worried that our notion of generalised causal structure is problematic since it relies on non-trivial connections in spacetime via wormholes, and those connections are not causal curves. Indeed, since wormholes, as we have seen in section \S\ref{3.3}, are typically non-traversable, i.e. there cannot be a causal curve through the wormhole (though there are exceptions, see \cite{maldacena2018traversable}), no signal can be sent from one side to the other of the wormhole. \\
However, note that all a relation needs to be admitted in the building blocks of the Humean's metaphysics is that it is a spatiotemporal relation. Moreover, as we showed in \S\ref{2}, these relations should be part of the causal structure of spacetime, supporting robust counterfactual correlations. For example, standard spacelike connections in Minkowski spacetime are not useful for the Humean since they do not support any robust counterfactual correlations. The relation $R_{ERB}$ of \textit{being connected by a wormhole with such and such interior geometry} is a spacetime relation too, as we argued in section \S\ref{4.1}. However, contrary to spacelike connections in Minkowski, the interior geometry of the ERB encodes entanglement properties, captured in the generalised causal structure by the requirement that $R_{ERB}$ be non-trivial. These two properties of $R_{ERB}$ allow the Humean to resolve her conflict with entanglement. \\
Furthermore, the fact that the wormholes are non-traversable is an essential feature of the proposal. If this were not the case, then one would be able to send signals through the wormhole, and thus, given that by the conjecture this should be equivalent to an entangled system, via the entanglement relation. However, the no signalling theorem of QM prohibits such signal transmission via entanglement since it would amount to superluminal information transmission. The fact that wormholes are non-traversable is a feature, not a bug of the proposal. In conclusion, not any spacetime relation can resolve the conflict between Humeanism and entanglement, only those of ER=EPR captured by $R_{ERB}$ can.


\section{Conclusions}\label{6}
In this paper, we have argued that the ER=EPR conjecture can be usefully employed to resolve the apparent contradiction between the metaphysical position of Humeanism and the relation of quantum entanglement. Moreover, we have shown that the notion of causal structure $\textbf{(CS)}$ can be useful in elucidating the conceptual structure of ER=EPR.\footnote{In particular, the notion of causal structure $\textbf{(CS)}$ usefully clarifies how ER=EPR solves the firewall paradox, which we discuss in \cite{cinti2020}.} Of course, our reliance on such a conjecture brings with itself a certain degree of risk for the conclusion that we draw. Independently of whether or not our argument ultimately succeeds in its goals, it might be the case that ultimately it is the world to show us wrong, by rendering the ER=EPR conjecture false. If this ended up being the case, then the Humean would need to find new avenues to defend her position. We do not claim to have proven Humeanism to be the best metaphysics of the world available. Indeed, we do not even want to claim to have definitively resolved its tensions with entanglement. Instead, in the spirit of naturalised metaphysics, we provided a possible, empirically falsifiable, avenue to connect the metaphysical hypothesis of Humeanism with some of the fascinating aspects of recent research in QG.

\section*{acknowledgements}
For helpful comment on an earlier draft we would like to thank Alberto Corti, Vincenzo Fano and Christian W\"uthrich, and two anonymous referees.

\bibliographystyle{chicago}

\bibliography{Bibliography}

\newcommand{\noop}[1]{}
\begin{thebibliography}{}

\bibitem[\protect\citeauthoryear{Almheiri, Engelhardt, Marolf, and
  Maxfield}{Almheiri et~al.}{2019}]{Almheiri_2019}
Almheiri, A., N.~Engelhardt, D.~Marolf, and H.~Maxfield (2019, Dec).
\newblock The entropy of bulk quantum fields and the entanglement wedge of an
  evaporating black hole.
\newblock {\em Journal of High Energy Physics\/}~{\em 2019\/}(12).

\bibitem[\protect\citeauthoryear{Almheiri, Hartman, Maldacena, Shaghoulian, and
  Tajdini}{Almheiri et~al.}{2020a}]{almheiri2020entropy}
Almheiri, A., T.~Hartman, J.~Maldacena, E.~Shaghoulian, and A.~Tajdini (2020a).
\newblock The entropy of hawking radiation.

\bibitem[\protect\citeauthoryear{Almheiri, Hartman, Maldacena, Shaghoulian, and
  Tajdini}{Almheiri et~al.}{2020b}]{Almheiri_2020}
Almheiri, A., T.~Hartman, J.~Maldacena, E.~Shaghoulian, and A.~Tajdini (2020b,
  May).
\newblock Replica wormholes and the entropy of hawking radiation.
\newblock {\em Journal of High Energy Physics\/}~{\em 2020\/}(5).

\bibitem[\protect\citeauthoryear{Almheiri, Mahajan, Maldacena, and
  Zhao}{Almheiri et~al.}{2020}]{Almheiri_2020b}
Almheiri, A., R.~Mahajan, J.~Maldacena, and Y.~Zhao (2020, Mar).
\newblock The page curve of hawking radiation from semiclassical geometry.
\newblock {\em Journal of High Energy Physics\/}~{\em 2020\/}(3).

\bibitem[\protect\citeauthoryear{Almheiri, Marolf, Polchinski, and
  Sully}{Almheiri et~al.}{2013}]{Almheiri_2013}
Almheiri, A., D.~Marolf, J.~Polchinski, and J.~Sully (2013).
\newblock {Black Holes: Complementarity or Firewalls?}
\newblock {\em JHEP\/}~{\em 02}, 062.

\bibitem[\protect\citeauthoryear{Ammon and Erdmenger}{Ammon and
  Erdmenger}{2015}]{Ammon:2015wua}
Ammon, M. and J.~Erdmenger (2015, 4).
\newblock {\em {Gauge/gravity duality}: {Foundations and applications}}.
\newblock Cambridge: Cambridge University Press.

\bibitem[\protect\citeauthoryear{Arntzenius}{Arntzenius}{2002}]{pittphilsci1051}
Arntzenius, F. (2002).
\newblock Is quantum mechanics pointless?

\bibitem[\protect\citeauthoryear{Baker}{Baker}{2020}]{baker2020knox}
Baker, D.~J. (2020).
\newblock Knox's inertial spacetime functionalism (and a better alternative).
\newblock {\em Synthese\/}, 1--22.

\bibitem[\protect\citeauthoryear{Baker, Halvorson, and Swanson}{Baker
  et~al.}{2015}]{Baker:2015gmy}
Baker, D.~J., H.~Halvorson, and N.~Swanson (2015).
\newblock {The Conventionality of Parastatistics}.
\newblock {\em Brit. J. Phil. Sci.\/}~{\em 66\/}(4), 929--976.

\bibitem[\protect\citeauthoryear{Bhogal and Perry}{Bhogal and
  Perry}{2017}]{Bhogal2017-BHOWTH}
Bhogal, H. and Z.~Perry (2017).
\newblock What the humean should say about entanglement.
\newblock {\em No\^us\/}~{\em 51\/}(1), 74--94.

\bibitem[\protect\citeauthoryear{Brown, Gharibyan, Leichenauer, Lin, Nezami,
  Salton, Susskind, Swingle, and Walter}{Brown et~al.}{2021}]{brown2021quantum}
Brown, A.~R., H.~Gharibyan, S.~Leichenauer, H.~W. Lin, S.~Nezami, G.~Salton,
  L.~Susskind, B.~Swingle, and M.~Walter (2021).
\newblock Quantum gravity in the lab: Teleportation by size and traversable
  wormholes.

\bibitem[\protect\citeauthoryear{Brown}{Brown}{2005}]{brown2005physical}
Brown, H.~R. (2005).
\newblock {\em Physical relativity: Space-time structure from a dynamical
  perspective}.
\newblock Oxford University Press on Demand.

\bibitem[\protect\citeauthoryear{Butterfield}{Butterfield}{2005}]{butterfield2005pointillisme}
Butterfield, J. (2005).
\newblock Against pointillisme about geometry.

\bibitem[\protect\citeauthoryear{Butterfield}{Butterfield}{2006}]{Butterfield2006-BUTAPA}
Butterfield, J. (2006).
\newblock Against pointillisme about mechanics.
\newblock {\em British Journal for the Philosophy of Science\/}~{\em 57\/}(4),
  709--753.

\bibitem[\protect\citeauthoryear{Butterfield and Gomes}{Butterfield and
  Gomes}{2020}]{butterfield2020functionalism}
Butterfield, J. and H.~Gomes (2020).
\newblock Functionalism as a species of reduction.
\newblock {\em arXiv preprint arXiv:2008.13366\/}.

\bibitem[\protect\citeauthoryear{Callender}{Callender}{2015}]{10.2307/24704648}
Callender, C. (2015).
\newblock One world, one beable.
\newblock {\em Synthese\/}~{\em 192\/}(10), 3153--3177.

\bibitem[\protect\citeauthoryear{Chirco, Goe{\ss}mann, Oriti, and Zhang}{Chirco
  et~al.}{2019}]{chirco2019group}
Chirco, G., A.~Goe{\ss}mann, D.~Oriti, and M.~Zhang (2019).
\newblock Group field theory and holographic tensor networks: Dynamical
  corrections to the ryu-takayanagi formula.

\bibitem[\protect\citeauthoryear{Cinti and Sanchioni}{Cinti and
  Sanchioni}{2021}]{cinti2020}
Cinti, E. and M.~Sanchioni (2021).
\newblock The devil in the (implicit) details: On the amps paradox and its
  resolution.

\bibitem[\protect\citeauthoryear{Dai, Minic, Stojkovic, and Fu}{Dai
  et~al.}{2020}]{PhysRevD.102.066004}
Dai, D.-C., D.~Minic, D.~Stojkovic, and C.~Fu (2020, Sep).
\newblock Testing the $\mathrm{ER}=\mathrm{EPR}$ conjecture.
\newblock {\em Phys. Rev. D\/}~{\em 102}, 066004.

\bibitem[\protect\citeauthoryear{Darby}{Darby}{2012}]{Darby2012-DARRHA}
Darby, G. (2012).
\newblock Relational holism and humean supervenience.
\newblock {\em British Journal for the Philosophy of Science\/}~{\em 63\/}(4),
  773--788.

\bibitem[\protect\citeauthoryear{D{\"u}rr, Goldstein, and Zangh{\`\i}}{D{\"u}rr
  et~al.}{2013}]{Drr2013}
D{\"u}rr, D., S.~Goldstein, and N.~Zangh{\`\i} (2013).
\newblock {\em Quantum Spacetime without Observers: Ontological Clarity and the
  Conceptual Foundations of Quantum Gravity}, pp.\  247--261.
\newblock Berlin, Heidelberg: Springer Berlin Heidelberg.

\bibitem[\protect\citeauthoryear{Einstein, Podolsky, and Rosen}{Einstein
  et~al.}{1935}]{PhysRev.47.777}
Einstein, A., B.~Podolsky, and N.~Rosen (1935, May).
\newblock Can quantum-mechanical description of physical reality be considered
  complete?
\newblock {\em Phys. Rev.\/}~{\em 47}, 777--780.

\bibitem[\protect\citeauthoryear{Einstein and Rosen}{Einstein and
  Rosen}{1935}]{PhysRev.48.73}
Einstein, A. and N.~Rosen (1935, Jul).
\newblock The particle problem in the general theory of relativity.
\newblock {\em Phys. Rev.\/}~{\em 48}, 73--77.

\bibitem[\protect\citeauthoryear{Esfeld}{Esfeld}{2014}]{Esfeld2014-ESFQHO}
Esfeld, M. (2014).
\newblock Quantum humeanism, or: Physicalism without properties.
\newblock {\em Philosophical Quarterly\/}~{\em 64\/}(256), 453--470.

\bibitem[\protect\citeauthoryear{Esfeld, Hubert, Lazarovici, and
  D{\"u}rr}{Esfeld et~al.}{2013}]{10.1093/bjps/axt019}
Esfeld, M., M.~Hubert, D.~Lazarovici, and D.~D{\"u}rr (2013, 09).
\newblock {The Ontology of Bohmian Mechanics}.
\newblock {\em The British Journal for the Philosophy of Science\/}~{\em
  65\/}(4), 773--796.

\bibitem[\protect\citeauthoryear{Fletcher, Manchak, Schneider, and
  Weatherall}{Fletcher et~al.}{2018}]{fletcher2018would}
Fletcher, S.~C., J.~B. Manchak, M.~D. Schneider, and J.~O. Weatherall (2018).
\newblock Would two dimensions be world enough for spacetime?
\newblock {\em Studies in History and Philosophy of Science Part B: Studies in
  History and Philosophy of Modern Physics\/}~{\em 63}, 100--113.

\bibitem[\protect\citeauthoryear{Glick and Darby}{Glick and
  Darby}{2018}]{pittphilsci14945}
Glick, D. and G.~Darby (2018, August).
\newblock In defense of the metaphysics of entanglement.

\bibitem[\protect\citeauthoryear{Halvorson}{Halvorson}{2007}]{halvorson2007algebraic}
Halvorson, H. (2007).
\newblock Algebraic quantum field theory.
\newblock In J.~Earman and J.~Butterfield (Eds.), {\em Handbook of Philosophy
  of Physics}, pp.\  731--922. Elsevier.

\bibitem[\protect\citeauthoryear{Han and Hung}{Han and Hung}{2017}]{Han_2017}
Han, M. and L.-Y. Hung (2017).
\newblock {Loop Quantum Gravity, Exact Holographic Mapping, and Holographic
  Entanglement Entropy}.
\newblock {\em Phys. Rev. D\/}~{\em 95\/}(2), 024011.

\bibitem[\protect\citeauthoryear{Harlow}{Harlow}{2016}]{Harlow:2014yka}
Harlow, D. (2016).
\newblock {Jerusalem Lectures on Black Holes and Quantum Information}.
\newblock {\em Rev. Mod. Phys.\/}~{\em 88}, 015002.

\bibitem[\protect\citeauthoryear{Healey}{Healey}{2016}]{healey2016pragmatist}
Healey, R. (2016).
\newblock A pragmatist view of the metaphysics of entanglement.
\newblock {\em Synthese\/}, 1--38.

\bibitem[\protect\citeauthoryear{Jackiw}{Jackiw}{1985}]{JACKIW1985343}
Jackiw, R. (1985).
\newblock Lower dimensional gravity.
\newblock {\em Nuclear Physics B\/}~{\em 252}, 343--356.

\bibitem[\protect\citeauthoryear{Jaksland}{Jaksland}{2020}]{pittphilsci17099}
Jaksland, R. (2020).
\newblock Entanglement as the world-making relation: Distance from
  entanglement.

\bibitem[\protect\citeauthoryear{Knox}{Knox}{2013}]{knox2013effective}
Knox, E. (2013).
\newblock Effective spacetime geometry.
\newblock {\em Studies in History and Philosophy of Science Part B: Studies in
  History and Philosophy of Modern Physics\/}~{\em 44\/}(3), 346--356.

\bibitem[\protect\citeauthoryear{Knox}{Knox}{2019}]{knox2019physical}
Knox, E. (2019).
\newblock Physical relativity from a functionalist perspective.
\newblock {\em Studies in History and Philosophy of Science Part B: Studies in
  History and Philosophy of Modern Physics\/}~{\em 67}, 118--124.

\bibitem[\protect\citeauthoryear{Ladyman, Ross, Collier, Spurrett, Spurrett,
  Collier, et~al.}{Ladyman et~al.}{2007}]{ladyman2007every}
Ladyman, J., D.~Ross, J.~Collier, D.~Spurrett, D.~Spurrett, J.~G. Collier,
  et~al. (2007).
\newblock {\em Every thing must go: Metaphysics naturalized}.
\newblock Oxford University Press on Demand.

\bibitem[\protect\citeauthoryear{Lam and W{\"u}thrich}{Lam and
  W{\"u}thrich}{2020}]{lam2020spacetime}
Lam, V. and C.~W{\"u}thrich (2020).
\newblock Spacetime functionalism from a realist perspective.
\newblock {\em Synthese\/}, 1--19.

\bibitem[\protect\citeauthoryear{Le~Bihan}{Le~Bihan}{2019}]{le2019spacetime}
Le~Bihan, B. (2019).
\newblock Spacetime emergence in quantum gravity: functionalism and the hard
  problem.
\newblock {\em Synthese\/}.

\bibitem[\protect\citeauthoryear{Le~Bihan}{Le~Bihan}{2020}]{le2020string}
Le~Bihan, B. (2020).
\newblock String theory, loop quantum gravity and eternalism.
\newblock {\em European Journal for Philosophy of Science\/}~{\em 10\/}(2),
  1--22.

\bibitem[\protect\citeauthoryear{Lewis}{Lewis}{1986}]{Lewis1986-LEWPPV-4}
Lewis, D. (1986).
\newblock {\em Philosophical Papers, Volume II}.
\newblock Oxford University Press.

\bibitem[\protect\citeauthoryear{Lyre}{Lyre}{2009}]{Lyre2010}
Lyre, H. (2009, April).
\newblock Humean perspectives on structural realism.

\bibitem[\protect\citeauthoryear{Malament}{Malament}{2012}]{Malament2012-MALTIT}
Malament, D.~B. (2012).
\newblock {\em Topics in the Foundations of General Relativity and Newtonian
  Gravitation Theory}.
\newblock Chicago: Chicago University Press.

\bibitem[\protect\citeauthoryear{Maldacena, Milekhin, and Popov}{Maldacena
  et~al.}{2018}]{maldacena2018traversable}
Maldacena, J., A.~Milekhin, and F.~Popov (2018).
\newblock Traversable wormholes in four dimensions.

\bibitem[\protect\citeauthoryear{Maldacena and Susskind}{Maldacena and
  Susskind}{2013}]{Maldacena:2013xja}
Maldacena, J. and L.~Susskind (2013).
\newblock {Cool horizons for entangled black holes}.
\newblock {\em Fortsch. Phys.\/}~{\em 61}, 781--811.

\bibitem[\protect\citeauthoryear{Maldacena}{Maldacena}{1999}]{Maldacena:1997re}
Maldacena, J.~M. (1999).
\newblock {The Large N limit of superconformal field theories and
  supergravity}.
\newblock {\em Int. J. Theor. Phys.\/}~{\em 38}, 1113--1133.
\newblock [Adv. Theor. Math. Phys.2,231(1998)].

\bibitem[\protect\citeauthoryear{Matarese}{Matarese}{2019}]{Matarese:2019bxo}
Matarese, V. (2019).
\newblock {Loop Quantum Gravity: A New Threat to Humeanism? Part I: The Problem
  of Spacetime}.
\newblock {\em Found. Phys.\/}~{\em 49\/}(3), 232--259.

\bibitem[\protect\citeauthoryear{Maudlin}{Maudlin}{2002}]{Maudlin2002-MAUQNA}
Maudlin, T. (2002).
\newblock {\em Quantum Non-Locality and Relativity: Metaphysical Intimations of
  Modern Physics}.
\newblock Blackwell.

\bibitem[\protect\citeauthoryear{Maudlin}{Maudlin}{2007}]{Maudlin2007-MAUTMW}
Maudlin, T. (2007).
\newblock {\em The Metaphysics Within Physics}.
\newblock Oxford University Press.

\bibitem[\protect\citeauthoryear{Mermin}{Mermin}{1985}]{Mermin1985IsTM}
Mermin, N.~D. (1985).
\newblock Is the moon there when nobody looks? reality and the quantum theory.
\newblock {\em Physics Today\/}~{\em 38}, 38--47.

\bibitem[\protect\citeauthoryear{Miller}{Miller}{2014}]{Miller2014-MILQEB}
Miller, E. (2014).
\newblock Quantum entanglement, bohmian mechanics, and humean supervenience.
\newblock {\em Australasian Journal of Philosophy\/}~{\em 92\/}(3), 567--583.

\bibitem[\protect\citeauthoryear{Myrvold}{Myrvold}{2016}]{myrvold2015lessons}
Myrvold, W.~C. (2016).
\newblock Lessons of bell's theorem: Nonlocality, yes; action at a distance,
  not necessarily.
\newblock In S.~Gao and M.~Bell (Eds.), {\em Quantum Nonlocality and Reality --
  50 Years of Bell's Theorem}. Cambridge University Press.

\bibitem[\protect\citeauthoryear{Nezami, Lin, Brown, Gharibyan, Leichenauer,
  Salton, Susskind, Swingle, and Walter}{Nezami
  et~al.}{2021}]{nezami2021quantum}
Nezami, S., H.~W. Lin, A.~R. Brown, H.~Gharibyan, S.~Leichenauer, G.~Salton,
  L.~Susskind, B.~Swingle, and M.~Walter (2021).
\newblock Quantum gravity in the lab: Teleportation by size and traversable
  wormholes, part ii.

\bibitem[\protect\citeauthoryear{Olson and Ralph}{Olson and
  Ralph}{2012}]{Olson:2011bq}
Olson, S. and T.~C. Ralph (2012).
\newblock {Extraction of timelike entanglement from the quantum vacuum}.
\newblock {\em Phys. Rev. A\/}~{\em 85}, 012306.

\bibitem[\protect\citeauthoryear{Papadodimas and Raju}{Papadodimas and
  Raju}{2013}]{Papadodimas:2012aq}
Papadodimas, K. and S.~Raju (2013).
\newblock {An Infalling Observer in AdS/CFT}.
\newblock {\em JHEP\/}~{\em 10}, 212.

\bibitem[\protect\citeauthoryear{Penington}{Penington}{2019}]{penington2019entanglement}
Penington, G. (2019).
\newblock Entanglement wedge reconstruction and the information paradox.

\bibitem[\protect\citeauthoryear{Penington, Shenker, Stanford, and
  Yang}{Penington et~al.}{2019}]{penington2019replica}
Penington, G., S.~H. Shenker, D.~Stanford, and Z.~Yang (2019).
\newblock Replica wormholes and the black hole interior.

\bibitem[\protect\citeauthoryear{Read and Menon}{Read and
  Menon}{2019}]{read2019limitations}
Read, J. and T.~Menon (2019).
\newblock The limitations of inertial frame spacetime functionalism.
\newblock {\em Synthese\/}, 1--23.

\bibitem[\protect\citeauthoryear{Redhead}{Redhead}{1987}]{Redhead1987-REDINA}
Redhead, M. (1987).
\newblock {\em Incompleteness, Nonlocality, and Realism: A Prolegomenon to the
  Philosophy of Quantum Mechanics}.
\newblock Oxford University Press.

\bibitem[\protect\citeauthoryear{Saad, Shenker, and Stanford}{Saad
  et~al.}{2019}]{saad2019jt}
Saad, P., S.~H. Shenker, and D.~Stanford (2019).
\newblock Jt gravity as a matrix integral.

\bibitem[\protect\citeauthoryear{Sol{\'e} and Hoefer}{Sol{\'e} and
  Hoefer}{2019}]{Sol2019}
Sol{\'e}, A. and C.~Hoefer (2019).
\newblock The nomological interpretation of the wave function.
\newblock In A.~Cordero (Ed.), {\em Philosophers Look at Quantum Mechanics},
  pp.\  119--138. Cham: Springer International Publishing.

\bibitem[\protect\citeauthoryear{Susskind}{Susskind}{2016}]{Susskind:2016jjb}
Susskind, L. (2016).
\newblock {Copenhagen vs Everett, Teleportation, and ER=EPR}.
\newblock {\em Fortsch. Phys.\/}~{\em 64\/}(6-7), 551--564.

\bibitem[\protect\citeauthoryear{Teitelboim}{Teitelboim}{1983}]{TEITELBOIM198341}
Teitelboim, C. (1983).
\newblock Gravitation and hamiltonian structure in two spacetime dimensions.
\newblock {\em Physics Letters B\/}~{\em 126\/}(1), 41--45.

\bibitem[\protect\citeauthoryear{Vistarini}{Vistarini}{2019}]{vistarini2019emergence}
Vistarini, T. (2019).
\newblock {\em The Emergence of Spacetime in String Theory}.
\newblock Routledge.

\bibitem[\protect\citeauthoryear{Wald}{Wald}{1984}]{Wald:1984rg}
Wald, R.~M. (1984).
\newblock {\em {General Relativity}}.
\newblock Chicago, USA: Chicago Univ. Pr.

\bibitem[\protect\citeauthoryear{Wallace}{Wallace}{2012}]{wallace2012emergent}
Wallace, D. (2012).
\newblock {\em The emergent multiverse: Quantum theory according to the Everett
  interpretation}.
\newblock Oxford University Press.

\bibitem[\protect\citeauthoryear{Wallace}{Wallace}{2020}]{Wallace:2017wzs}
Wallace, D. (2020).
\newblock {Why Black Hole Information Loss is Paradoxical}.
\newblock In N.~Huggett, K.~Matsubara, and C.~W\"uthrich (Eds.), {\em Beyond
  Spacetime}. Cambridge University Press.

\bibitem[\protect\citeauthoryear{W\"uthrich}{W\"uthrich}{2019}]{pittphilsci15902}
W\"uthrich, C. (2019, April).
\newblock When the actual world is not even possible.
\newblock 20 pages, 1 figure; April 2019 revision; for George Darby, David
  Glick, and Anna Marmodoro (eds.), The Foundation of Reality: Fundamentality,
  Space and Time, Oxford University Press.

\end{thebibliography}

\end{document}